\documentclass{aa}
\usepackage{graphicx}
\usepackage{txfonts}
\usepackage{amssymb}
\usepackage{ulem}  

\begin{document}

\title{Turbulence transmission in parallel relativistic shocks
       using ray tracing}

\author{Joni Tammi \inst{1,2}
 \and  Rami Vainio \inst{3}}
\offprints{J.\ Tammi; \email{joni.tammi@iki.fi}}

\institute{%
  Tuorla Observatory, V\"ais\"al\"a Institute for Space Physics and Astronomy,
  V\"ais\"al\"antie 20, FI-21500 Piikki\"o, Finland
  \and
  UCD School of Mathematical Sciences, University College Dublin,
  Belfield, Dublin 4, Ireland
  \and
  Department of Physical Sciences, P. O. Box 64, 
  FI-00014 University of Helsinki, Finland
}

\date{Received 10.1.2006 / Accepted 6.9.2006}

\abstract{ 
  We apply a semi-classical approach of handling waves as
  quasiparticle gas in a slowly varying flow -- analogous to
  ray tracing -- to calculate the Alfv\'en wave transmission
  parameters, the resulting cross-helicity of the waves and the
  scattering-centre compression ratio, for cases where the shock
  thickness is large enough for the turbulent waves in the plasma to
  see the transition of the background flow parameters as smooth and
  slowly varying. For nonrelativistic shocks the wave transmission
  produces similar effects on the downstream turbulence and the
  scattering-centre compression ratio as does the transmission through
  a step shock: the downstream Alfv\'en waves propagate predominantly
  towards the shock in the local plasma frame and, thus, the
  scattering-centre compression ratio is larger than the gas
  compression ratio. For thick relativistic shocks, however, we find
  qualitative differences with respect to the step-shock case: for
  low-Alfv\'enic-Mach-number shocks the downstream waves propagate
  predominantly away from the shock, and the scattering-centre
  compression ratio is lower than that of the gas.  Thus, when taken
  into account, the Alfv\'en wave transmission can decrease the
  efficiency of the first-order Fermi acceleration in a thick
  relativistic shock.
 }

\authorrunning{Tammi \& Vainio}
\titlerunning{Turbulence transmission in parallel shocks}

\maketitle

%
%
%
%

\section{Introduction}

A common approximation in shock-related particle acceleration studies
involving the first-order Fermi mechanism is to assume the scattering
turbulence frozen-in to the plasma. This is very well justified in
cases where the speed of the turbulent waves is negligibly small
compared to the flow speed. For cases where the wave speed increases
and becomes a notable fraction of the flow speed, however, the
deviation of the scattering centre speed from the speed of the
background flow can have notable effects on the particle acceleration
scenario; it was noted already by Bell (\cite{Bell1978}) that the
speed of the scattering centres is not necessarily that of the flow.
Furthermore, the compression ratio (of the scattering centres) is
known to have significant effect on the power-law spectral index of
particles accelerated in the shock for both steplike and thick shocks
(for modified shocks see Drury et al. \cite{DruryEtAl1982} and Drury
\cite{Drury1983} for nonrelativistic analytical studies, and Virtanen
\& Vainio 2005a, hereafter \cite{VV2005AA}%
\footnote{In Eq.~(1) of VV05 there is a typo in the exponent in the
denominator of the last term: $r^2$ should be $r^3$.}%
, for numerical results for all speeds, and Keshet \& Waxman
\cite{KeshetWaxman2005} for an analytical solution for step
shocks). The turbulent waves are, of course, also affected by the
shock, and thus one has to start from the transmission of turbulence
through the shock to arrive at describing particle acceleration in a
case where wave-propagation effects are taken into account.

The effect of the transmission of Alfv\'en waves through a step shock
was solved at the nonrelativistic limit by Vainio \& Schlickeiser
(\cite{VS1998}) and later generalised to include also relativistic
shocks by Vainio et al.\ (2003, hereafter referred to as
\cite{VVS2003}, and \cite{VVS2005}). They found that, regardless of
the shock speed, waves initially in equipartition were propagating
predominantly antiparallel to the flow direction (i.e.\ backwards
toward the shock in the downstream plasma frame) after being
transmitted through a step shock with Alfv\'enic Mach number
sufficiently low to yield non-negligible wave speeds compared to the
speed of the plasma flow. They showed that this leads to an increased
\textit{scattering-centre compression ratio}, i.e., that the
scattering-centre speed can undergo compression that is significantly
higher than the compression of the flow itself. This effect thus
causes the first-order Fermi acceleration to lead to accelerated
particle energy spectral indices harder than $\sim 2.2$, the
well-known outcome of parallel step shocks at the ultrarelativistic
limit in case of isotropic particle scattering (e.g., Kirk \& Duffy
\cite{KD1999}, Keshet \& Waxman \cite{KeshetWaxman2005}). This
transmission analysis is, however, valid only for shocks of small
thickness and waves with wavelengths much longer than the width of the
shock transition.

In this paper we continue studying the transmission of Alfv\'en waves
through parallel shocks, but take an approach opposite to the
step-shock approximation of \cite{VVS2003}. We calculate the
transmission coefficients and their derivative parameters (the average
cross-helicity of the waves, and the compression ratio of the
scattering centres) for a continuous flow profile, approximating the
transmission of waves through a shock with thickness sufficiently
large for the waves to see the flow parameters varying slowly. The
transmission coefficients for the nonrelativistic thick-shock case
were calculated by \cite{VV2005AA}. 

In general, a single shock wave is not either a step-shock nor a thick
shock: as the turbulence in the upstream medium has a broad-band
spectrum, some waves will see the shock as a thick while others see it
as a thin structure. In addition, some waves have wavelengths matching
the shock thickness. Thus, the general theory of wave propagation can
not rely on either the thick or thin shock approximation. However, the
general theory is mathematically very cumbersome and we have here
chosen to limit our discussion to waves with wavelengths much shorter
than the shock thickness. Combining these results with our earlier
ones (\cite{VVS2003}) with the opposite assumption gives the first
qualitative view on turbulence transmission through parallel shocks
with finite thickness.

%
%
%
%

\section{Alfv\'en wave transmission}


\subsection{Theoretical background}

We adopt the idea of treating the waves as a quasiparticle gas. For
monochromatic waves, the number density of the wave quanta, or the
wave action density, is $N' = E'/\omega'$, where $E'$ and $\omega'$
are the energy density and the angular frequency of the waves in the
plasma frame. We will use the prime to denote the quantities measured
in the plasma frame, while unprimed values are measured in the shock
frame. Below, we will consider a broad-band spectrum of waves and,
thus, the evolution of the quasi-particle distribution function in
the phase space $(\mathbf{x},\mathbf{k})$.

Our approach is equivalent to the use of a variational principle in
deriving the equation for the turbulent MHD wave-action density,
pioneered by Dewar (\cite{Dewar1970}) and extended to the relativistic
case by Achterberg (\cite{Achterberg1983}).

In addition to the angular frequency $\omega$, we will be needing the
wavenumber $k$ and frequency $f=\frac{\omega}{2\pi}$, as well as their
values in the local plasma frame ($\omega'$, $k'$ and $f' =
\frac{\omega'}{2\pi}$ respectively). Because the medium consists of a
time-independent flow profile through the shock, the shock-frame
frequencies are conserved during shock crossing and we can write the
correspondence between the shock- and plasma-frame values as
\begin{equation} \label{eq:wavenumbers}
  k = \Gamma ( k' + V \omega' ) 
    = \pm \Gamma (1 \pm V V_{\rm A} ) \omega' / V_{\rm A}
\end{equation}
and
\begin{equation} \label{eq:frequencies}
  f = \Gamma (\omega' + V k') / (2\pi) 
    = \pm \Gamma (V \pm V_{\rm A}) f' / V_{\rm A},
\end{equation}
where the dispersion relation $\omega' = \pm V_{\rm A} k'$ is used. The
Alfv\'en speed $V_{\rm A}$ is obtained from the proper Alfv\'en speed (in
units of $c$) as
\begin{equation}
  u_{\rm A} = \Gamma_{\rm A} V_{\rm A} = B'_0 / \sqrt{4 \pi \mu' n'},
\end{equation}
where \( \mu' = (\rho' + P')/n' \), $n'$, $\rho'$ and $P'$ are the
specific enthalpy, the number density, the total energy density and
the gas pressure, measured in the local plasma frame, and $B'_0$ is 
the magnitude of the of the static large scale background magnetic field. 
For easier comparison with the step-shock transmission of 
\cite{VVS2003} we will present results using the proper speed 
$ u = \Gamma V $, where \( \Gamma = 1 / \sqrt{1 -V^2} \) is the 
Lorentz factor, and the speeds are measured in units of $c$. 
Using subscripts $1$ and $2$ for far upstream and 
downstream values, respectively, we can introduce the Alfv\'enic Mach 
number of the shock \( M^2 = u_1^2 / u_{\rm A,1}^2 \). 

We adopt the turbulence spectra in a power-law form
\begin{equation}
  I'^\pm(k',x) = I'^\pm_0 \cdot (k'_0 / k' )^q \quad {\rm for} 
  \quad  k' > k'_0,
\end{equation}
at wavelengths greater than $k'_0$, for waves flowing parallel 
(also referred to as forward waves, denoted by ''$+$'') and antiparallel 
(or backward, denoted by ''$-$'') to the flow direction.

%
\subsection{Equation for the wave flux}
\label{sec:waveflux}

Let us consider the propagation of an Alfv\'en wave in a
time-independent medium with spatial gradients only in the direction
of the background flow, aligned with the background magnetic field and
the $x$-axis. The Hamiltonian of the system is
\begin{equation}
  H(\vec{x},\vec{p})=\hbar\Omega(x,p_{x}/\hbar),
\end{equation}
where $\vec{x}$ and $\vec{p}=\hbar\vec{k}$ are the canonical coordinates
and momenta, $\vec{k}$ is the wavevector, and 
\begin{equation}
  \omega=\Omega(x,k_{x})=V_{\rm g}(x)\, k_{x}
\end{equation}
gives the dispersion relation of the Alfv\'en wave. Here 
\mbox{$V_{\rm g} = (V + sV_{\rm A} ) / ( 1 + sVV_{\rm A} )$}
is the group speed of the wave and $s=\pm1$ gives its propagation
direction with respect to the background plasma flow. Hamilton's equations
of motion for the wave packet, thus, read
\begin{equation}
  \dot{x} = \frac{\partial H}{\partial p_{x}} 
  = \frac{\partial\Omega}{\partial k_{x}}; 
  \quad
  \dot{k}_{x} = \frac{\dot{p}_{x}}{\hbar} 
  = -\frac{1}{\hbar}\frac{\partial H}{\partial x}
  = -\frac{\partial\Omega}{\partial x},
\end{equation}
with all other equations of motion being trivial: 
$\dot{y}=\dot{z}=\dot{p}_{y}=\dot{p}_{z}=0$
because the Hamiltonian has no dependence on these phase space coordinates.
Thus, the continuity equation for the phase space density, 
\begin{displaymath}
  f(\vec{x},\vec{p}) = \frac{ {\rm d} N }{ {\rm d}^{3}x\, {\rm d}^{3}p }
  = \frac{1}{\hbar^{3}} \frac{{\rm d}N}{{\rm d}^{3}x\, {\rm d}^{3}k}
  \equiv \frac{1}{\hbar^{3}} N(\vec{x},\vec{k}),
\end{displaymath}
of the waves (or Liouville's equation, equivalently) yields
\begin{equation}
  \frac{\partial N}{\partial t} + \frac{\partial}{\partial\vec{x}} 
  \cdot \left(\dot{\vec{x}}N\right) + \frac{\partial}{\partial\vec{k}}
  \cdot \left(\dot{\vec{k}}N\right) = 0,
\end{equation}
which, after the substitution of the equations of motion, reduces to
\begin{equation}
  \frac{\partial N}{\partial t} + \frac{\partial}{\partial x}
  \left( \frac{\partial\Omega}{\partial k_{x}}N \right) - 
  \frac{\partial}{\partial k_{x}} 
  \left( \frac{\partial\Omega}{\partial x}N \right) = 0.
\end{equation}
We can integrate this equation over the perpendicular wave numbers
$k_{y}$ and $k_{z}$ to obtain an equation for the number-density
of waves over $k_{x}$, 
\begin{displaymath}
  n_{k_{x}}(\vec{x},k_{x}) 
  = \frac{ {\rm d} N }{ {\rm d}^{3}x\, {\rm d}k_{x} } 
  = \iint {\rm d}k_{y} \, {\rm d}k_{z}\, N(\vec{x},\vec{k}),
\end{displaymath}
as
\begin{equation}
  \frac{\partial n_{k_{x}}}{\partial t} + 
  \frac{\partial}{\partial x}\left(\frac{\partial\Omega}{\partial k_{x}}\, 
    n_{k_{x}}\right) -
  \frac{\partial}{\partial k_{x}}\left(\frac{\partial\Omega}{\partial x}\, 
    n_{k_{x}}\right)=0,
\end{equation}
because $\Omega$ does not depend on $k_{y}$ or $k_{z}$. This equation
holds regardless of the form of the distribution function with respect
to the wavenumbers $k_{y}$ and $k_{z}$. These are, of course, constants
of motion for the waves in our system.

From now on, we will assume that the wavevector is aligned with the
$x$-axis (i.e., that $k_{y}=k_{z}=0$) and simplify the notation
by writing $k_{x}=k$. Thus, using the explicit form of the dispersion
relation, we obtain in steady state
\begin{equation}
  \frac{\partial}{\partial x}\left(V_{\rm g}\, n_{k}\right)-\frac{\partial}{\partial k}\left(\frac{\partial V_{\rm g}}{\partial x}k\, n_{k}\right)=0.
\end{equation}
Consider, next, the wave-action density over frequency, 
\mbox{$ n_{\omega} = {\rm d} N / ({\rm d}^{3}x\, {\rm d}\omega)=n_{k}/V_{\rm g}$,}
and change to using the position $x$ and the wave frequency 
$\omega=V_{\rm g}(x)\, k$
as the independent variables instead of $x$ and $k$. Thus,
\begin{eqnarray*}
\left(\frac{\partial}{\partial x}\right)_{k} & \to & \left(\frac{\partial}{\partial x}\right)_{\omega}+\frac{dV_{{\rm g}}}{dx}\frac{\omega}{V_{{\rm g}}}\left(\frac{\partial}{\partial\omega}\right)_{x}\\
\left(\frac{\partial}{\partial k}\right)_{x} & \to & V_{{\rm g}}\left(\frac{\partial}{\partial\omega}\right)_{x}\end{eqnarray*}
and we obtain\begin{eqnarray*}
0 & = & \left[\left(\frac{\partial}{\partial x}\right)_{\omega}+\frac{dV_{{\rm g}}}{dx}\frac{\omega}{V_{{\rm g}}}\left(\frac{\partial}{\partial\omega}\right)_{x}\right](V_{{\rm g}}\cdot V_{{\rm g}}n_{\omega})\\
 &  & \quad-V_{{\rm g}}\left(\frac{\partial}{\partial\omega}\right)_{x}\left(\frac{dV_{{\rm g}}}{dx}\omega\, n_{\omega}\right)\\
 & = & V_{{\rm g}}\left(\frac{\partial}{\partial x}(V_{{\rm g}}n_{\omega})\right)_{\omega}\end{eqnarray*}
In this equation $\omega$ appears only as a parameter numbering the
modes, which can be all treated as monochromatic waves using the constancy
of their flux
\begin{equation}
  V_{\rm g} n_{\omega} = n_{k}(x,\omega/V_{\rm g}(x))
\end{equation}
with respect to position at constant $\omega$. Thus,
\begin{displaymath}
  n_{k}(x,k) = n_{k}(-\infty,k_{1}),
\end{displaymath}
where $k_{1}(k,x) = V_{\rm g}(x)k / V_{\rm g,1}$ is the far-upstream wavenumber
of the wave with wavenumber $k$ at $x$.

%
\subsection{Wave transmission coefficients}

Next we calculate the transmission coefficients for Alfv\'en waves with
wavelengths much shorter than the shock thickness. 

Transmission coefficients are needed for solving the turbulence
conditions at a given location in the shock. In contrast to the step
shock case (\cite{VVS2003}) where part of the waves are reflected at
the shock and change their mode (from parallel to antiparallel, or
vice versa), in the case of slowly changing medium all the waves are
transmitted through the shock without reflection, and the turbulence
at a given location can be obtained from equation
\begin{equation}
  I'^\pm(k',x) = T_\pm'^2(x) \cdot I'^\pm_1(k'), 
\end{equation}
where \( I'^\pm_1(k') = I'^\pm(k',x=-\infty) \) are the turbulence
spectra far upstream for parallel ($+$) and antiparallel ($-$) waves,
and $T'^2_\pm(x)$ are the corresponding transmission coefficients, now
to be solved.

The co-moving intensity of the waves is related to the number-density
of wave quanta by
\begin{eqnarray*}
  I'^{\pm}(x,k') & = & \frac{\hbar\omega'\,    {n'_{k'}}^{\pm}   }{4\pi}
  = \frac{V_{{\rm A}}k'}{V_{\rm g}^{\pm}k}\frac{\hbar\omega\, n_{k}^{\pm}}{4\pi}\\
  & = & \frac{V_{{\rm A}}}{V_{\rm g}^{\pm}\Gamma(1\pm VV_{{\rm A}})}
  \frac{\hbar\omega\, n_{k}^{\pm}}{4\pi}
  = \frac{V_{{\rm A}}}{\Gamma(V\pm V_{{\rm A}})}
  \frac{\hbar\omega\, n_{k}^{\pm}}{4\pi},
\end{eqnarray*}
where we have made use of the invariance of $n_{k}^{\pm}$ under
Lorentz transformations along the $x$ axis. (This holds because of the
Lorentz invariance of $N(\vec{x},\vec{k})$ and of the perpendicular
wavenumber element ${\rm d}k_{y}\, {\rm d}k_{z}={\rm d}k_{y}'\, {\rm
  d}k_{z}'$.) Thus, we get
\begin{displaymath}
  \frac{\Gamma(V\pm V_{{\rm A}})}{V_{{\rm A}}}\,{I'}^{\pm}(x,k')
  = \frac{\hbar\omega\, n_{k}^{\pm}}{4\pi}
  = \frac{\Gamma_{1}(V_{1}\pm V_{{\rm A}1})}{V_{{\rm A}1}}\,{I'_{1}}^{\pm}(k_{1}'),
\end{displaymath}
where $\Gamma_{1}(V_{1}\pm V_{{\rm A,1}})k'_{1}=\omega=\Gamma(V\pm V_{{\rm A}})k'$.
For power-law spectra, ${I'_{1}}^{\pm}(k')\propto k'^{-q}$, we finally
obtain
\begin{equation}
  {I'}^{\pm}(x,k') 
  = {I'_{1}}^{\pm}(k')\frac{V_{{\rm A}}}{V_{{\rm A,1}}}
  \left[
    \frac{\Gamma_{1}(V_{1}\pm V_{{\rm A}1})}{\Gamma(V\pm V_{{\rm A}})}
  \right]^{q+1}.
\end{equation}

At the nonrelativistic limit in the far downstream, 
this reduces to (\cite{VV2005AA})
\begin{equation} \label{eq:I2_nonrelativistic}
  I_2^\pm(k') =  I_1^\pm(k') \cdot T'^2_\pm = I_1^\pm(k') r^{q+1/2}
  \left[ \frac{ M\pm 1 }{ M \pm r^{1/2} } \right]^{q+1},
\end{equation}
where $r = V_1 / V_2$ is the gas compression ratio and $M$ reduces to
its nonrelativistic form, $M = V_1 / V_{{\rm A,}1}$.

%
%
%
%

\section{Results and discussion}

From this point on, we deal only with plasma-frame quantities, so we
omit the primes in variable symbols and assume them being measured in
the plasma rest frame, unless otherwise mentioned. The local plasma
speeds $V$ and the location are, of course, still in the shock frame.

For a general view we first look at the case of vanishing upstream
cross-helicity. The ratio of forward-to-backward waves in the
downstream is
\begin{equation}
  \frac{I^+}{I^-} = 
  \left[ 
   \frac{ (V_1+V_{\rm A,1})(V_2-V_{\rm A,2}) }
        { (V_1-V_{\rm A,1})(V_2+V_{\rm A,2}) }
  \right]^{q+1}
\end{equation}
which, in contrast to transmission in step shocks, can be also greater
than one (i.e., the downstream cross-helicity is positive) if $V_{\rm
  A,2} < V_{\rm A,1}/r$. This, however, can occur only if the shock is
relativistic (for nonrelativistic shock $V_{\rm A,2} = V_{\rm
  A,1}/\sqrt{r}$ and $I^+/I^- \le 1$ always). 
One immediately sees that, as a consequence, while
the effect of the transmission in the nonrelativistic regime is very
similar to that of both non- and fully relativistic step shocks, for
relativistic shocks qualitative differences arise.

In the following sections we study the cross-helicity of the
transmitted waves in the downstream, and its development throughout
the shock. We also calculate the scattering centre compression ratios
in the downstream.

%
\subsection{Far downstream cross-helicity}  \label{subsec:hc2}

\begin{figure}
\includegraphics[width=1.0\linewidth]{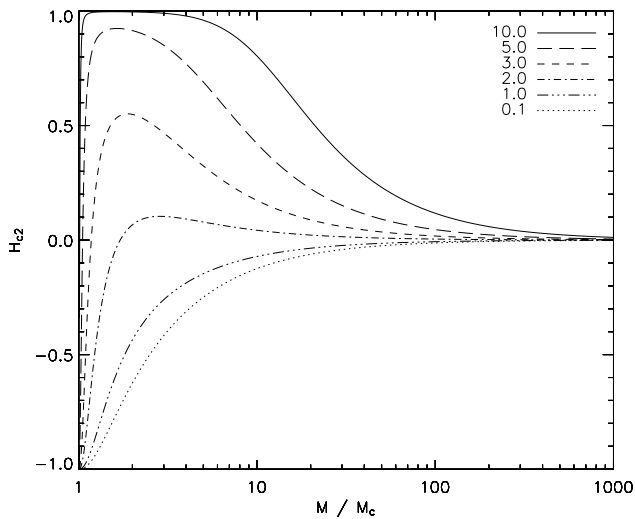}
\caption{The far downstream cross-helicity $H_{\rm c,2}$ as a function 
  of Alfv\'enic Mach number $M$ for different shock proper speeds $u_1
  = V_1 \Gamma_1$.  $M_{\rm c} = \sqrt{r}$ is the critical Mach
  number, below which the downstream Alfv\'en speed exceeds the local
  plasma speed, and the shock becomes non-evolutionary. The spectral
  index of the turbulence power-spectrum is $q=1.5$}
\label{fig:h_c2}
\end{figure}

Here we calculate the normalised cross-helicity of the waves, 
\begin{equation}
  H_{\rm c}(k,x) = \frac{I^+(k,x) - I^-(k,x)}{I^+(k,x) + I^-(k,x)},
\end{equation}
in the far downstream ($H_{\rm c,2}$) from a given upstream state 
($H_{\rm c,1}$) as a function of different Alfv\'enic Mach numbers.

For high-$M$ shocks, where the effect of wave speeds is negligible,
the downstream cross-helicity approaches, of course, the cross
helicity upstream as $M \to \infty$. As the Alfv\'enic Mach number
decreases and approaches the critical Mach number $M_{\rm c} =
\sqrt{r} $ (i.e., as the downstream Alfv\'en speed starts to approach
the downstream flow speed), the cross-helicity for nonrelativistic
shocks starts to decrease and approach $-1$ (as shown by
\cite{VV2005AA}), as for step shocks of all speeds (\cite{VVS2003}).

For relativistic speeds and Mach number a few times the critical
$M_{\rm c}$, however, nearly all of the waves are streaming
\textit{parallel} to the flow and $H_{\rm c,2} \to +1$ as $u_1 \to
\infty$ almost regardless of the upstream cross-helicity.  For the
lowest $M \gtrsim M_{\rm c}$ the cross-helicity drops to $-1$ as the
downstream Alfv\'en speed becomes equal to the flow speed; this
happens because the transmission coefficients of the backward waves,
$T^2_-$, goes to infinity. The physicality and removal of this
mathematical singularity is discussed in section
\ref{subsec:limitations}.

The lack of wave reflection causes the special cases of only one
upstream wave field to keep also the downstream wave field at the same
state (i.e., \( H_{\rm c,1} = \pm 1 \, \Rightarrow \, H_{\rm c,2} =
\pm 1 \)) but the above-mentioned general behaviour is seen for all 
\( |H_{\rm c,1}| < 1 \).  The downstream cross-helicity as a function of
the Alfv\'enic Mach number is plotted in Fig.~\ref{fig:h_c2} for
non-to-highly relativistic shocks for vanishing upstream cross
helicity.

%
\subsection{Cross-helicity as a function of location}

\begin{figure}
\includegraphics[width=1.0\linewidth]{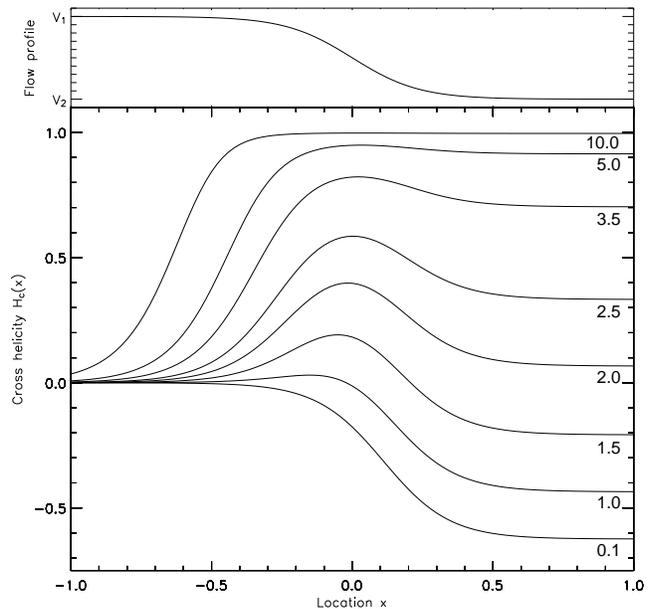}
\caption{ Form of the flow profile (upper panel) and the local cross
  helicity across the shock when $M/M_{\rm c} = 2$ (lower
  panel). Shock proper speeds $u_1$ range from $0.1\, c$ to $10\, c$
  (lines from bottom to top, respectively). In the upper panel the
  flow speed is in arbitrary units, showing only the form of the
  transition from far upstream speed $V_1$ to that of the downstream,
  $V_2$, and the spatial range across which the transition takes place
  for both the flow speed and the cross-helicities. See text for
  details.}
\label{fig:h_cx}
\end{figure}

Next we want to study how the cross-helicity changes from the upstream
value to that of the downstream. We use a hyperbolic tangent profile
(Schneider \& Kirk \cite{SK1989}, Virtanen \& Vainio \cite{VV2005ApJ})
to describe the flow speed as a function of location, and follow the
development of the local mean cross-helicity.  Since the actual shock
thickness is of no importance for the spatial properties of the
test-wave transmission, as long as it is large enough to allow for the
''smoothly-varying-background'' assumption, we use only one flow
profile (and one thickness) to demonstrate the general behaviour. The
form of the flow speed profile is shown in the top panel of
Fig.~\ref{fig:h_cx}. The unit of location $x$ in that figure is
arbitrary and is chosen so that the shock transition takes place
within $\approx 1$ location unit.

While for nonrelativistic shocks the cross-helicity changes
monotonically from the upstream value to that calculated in Section
\ref{subsec:hc2}, for faster shocks the local cross-helicity $H_{\rm
c} (x)$ has non-monotonic behaviour: the cross-helicity forms a
maximum near the ''centre'' of the flow profile, where the velocity
gradient has its largest value, after which it decreases to the
downstream value $H_{\rm c2}$.

An example of the non-monotonic cross-helicity is shown in
Fig.~\ref{fig:h_cx}, where the local cross-helicity $H_{\rm c} (x)$
(here set equal to zero the far upstream) is plotted as a function of
location $x$ for several shocks with speeds ranging from non- to fully
relativistic, but all having the Alfv\'enic Mach number $M = 2 M_{\rm
c}$. The critical Mach number $M_{\rm c} = \sqrt{r}$ is calculated
separately for each shock speed following the scheme described in
\cite{VVS2003}. The spectral index of the turbulence is $q=1.5$ for
all cases.

\subsection{Decreased compressio ratio}

\begin{figure}
  \includegraphics[width=1.0\linewidth]{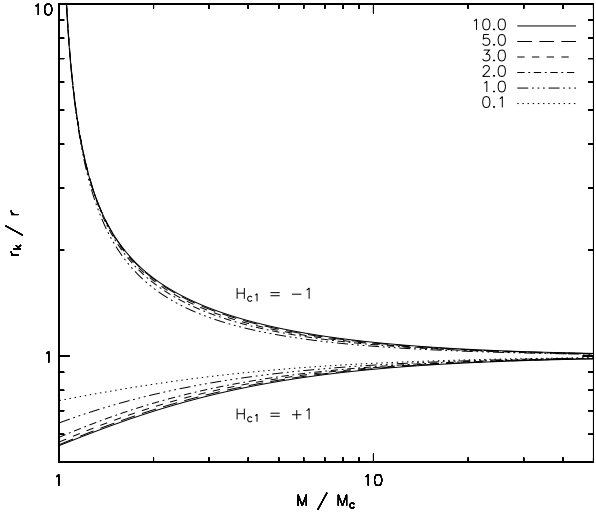}
  \caption{Ratio of the scattering centre compression ratio $r_{k}$ to
    that of the gas $r$ for different shock proper speeds $u_1$ as
    functions of the Alv\'enic Mach number $M$. Lines are drawn for
    upstream cross-helicity $H_{\rm c1} = -1$ (group of the uppermost
    lines) and $+1$ (lower group). The spectral index of the
    turbulence power spectrum is $q=1.5$ and $M_{\rm c}=\sqrt{r}$; see
    text for details.}
  \label{fig:rkr_hc1}
\end{figure}

While the assumption of small-amplitude waves leaves the compression
of the plasma itself unaffected, the compression felt by the
scattering particles, on the other hand, can be changed remarkably.
This is due to the fact that the particles are scattered by the
turbulent waves, and if the waves have speed non-neglibigle compared to
the speed of the flow (i.e. waves are not assumed to be frozen-in to
the plasma), the speed profile of the scattering centres can be
different from that of the plasma. Thus, also the effective
compression ratio felt by the particles, $r_k$, can
differ from the gas compression ratio $r$.
Using the local cross-helicity $H_{\rm c}$ we can express the
average wave speed as
\begin{equation} 
  V_{k}(x) = \frac{ V(x) + H_{\rm c}(x) V_{\rm A}(x) }
                   { 1   + H_{\rm c}(x) V(x) V_{\rm A}(x)}.
\end{equation}
Now the scattering-centre compression ratio is the calculated using
the average wave speeds ($V_{k,1}$ and $V_{k,2}$): 
\( r_k = V_{k,1} / V_{k,2} \).

As noted earlier, if the cross-helicity in the upstream equals $\pm 1$ 
it remains unchanged throughout the shock. So if there are e.g.\ only
forward waves in the upstream, the upstream scattering centre speed is
simply the (relativistic) sum of the flow speed and the local Alfv\'en
speed; and likewise in the downstream.  
However, because of the compression and drop in the flow speed at the
shock, the ratio of the Alfv\'en and the flow speed in the downstream
is larger than in the upstream. Also its effect on the resulting
shock-frame scattering-centre speed is larger when the underlying flow
speed is not close to the speed of light.
For only forward waves in the upstream this results in \( r_k < r \),
and for only backward waves in the upstream \( r_k > r \).  When the
Alfv\'en speed drops and becomes negligible compared to local flow
speeds (i.e. $M \to \infty$), \( r_k \to r \). This is illustrated in
Fig.~\ref{fig:rkr_hc1}.

For vanishing upstream cross-helicity, the behaviour in a
nonrelativistic shock was examined by \cite{VV2005AA} and it was shown
to be very similar to that in step shocks. Now for thick relativistic shock,
because of the positive downstream cross-helicity, the average
shock-frame wave speed in the downstream is higher than the local flow
speed, and, thus, the compression ratio of the scattering centres is
lower than that of the gas.
\begin{figure}
  \includegraphics[width=1.0\linewidth]{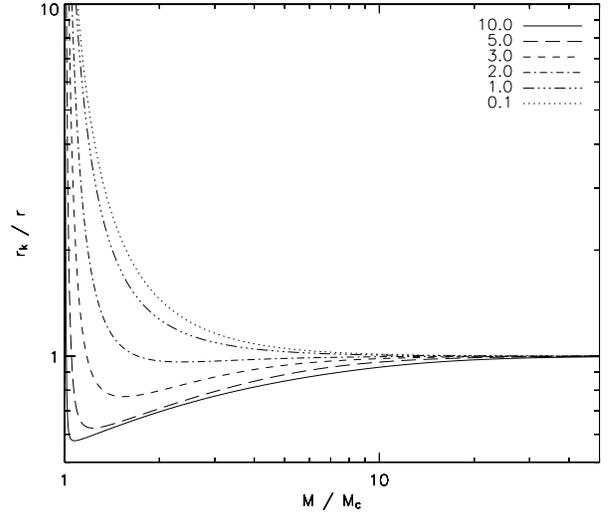}
  \caption{Same as Fig.~\ref{fig:rkr_hc1}, but for $H_{\rm c1} = -0$.}
  \label{fig:rkr_hc0}
\end{figure}

Resulting scattering-centre compression ratios (scaled to the gas
compression ratio) are plotted for upstream cross-helicities $H_{\rm
c,1} = \pm 1$ in Fig.~\ref{fig:rkr_hc1}, and for $H_{\rm c,1} = 0$ in
Fig.~\ref{fig:rkr_hc0}. The nonrelativistic ($u_1 = 0.1$) case is the
same as in \cite{VV2005AA} and, as for step shocks, the compression
ratio increases also for higher speeds.  However, \textit{on the
contrary} to what \cite{VV2005AA} expected, for the fastest low-$M$
shocks the scattering-centre compression ratio does not increase
steadily as $M \to M_{\rm c}$, but falls below the gas compression
ratio when the shock speed becomes relativistic.

\subsection{Limitations for the analysis} \label{subsec:limitations}

\begin{figure}
  \includegraphics[width=\linewidth]{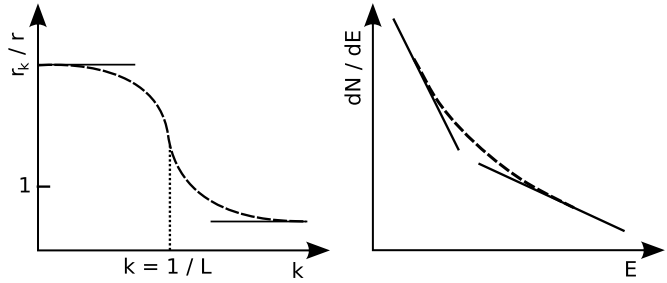}
  \caption{Sketch of the scattering-centre compression ratio as a
    function of wavenumber (left) and the resulting accelerated particle
    spectrum (right) in a thick relativistic shock. $L$ denotes the shock
    thickness.}
  \label{fig:f5}
\end{figure}

In earlier studies by Vainio \& Schlickeiser (\cite{VS1998}) for
nonrelativistic shocks and \cite{VVS2003} for relativistic ones, the
transmission properties were solved for waves of wavelengths
essentially longer than the length scales of the shock. Here we have
studied the opposite of ''short'' waves flowing through a shock
thicker than the wave length scales.  Although these two transmission
schemes work differently and in different parts of parameter space, it
would be physically meaningful to apply both schemes in the same
shock: the analysis of \cite{VVS2003} for the part of the turbulence
spectrum for which the wavelengths are longer than the width of the
shock transition, and the method described in this paper for waves for
which the plasma parameters vary slowly throughout the
transition. Thus, the scattering-centre compression ratio is a
function of wavenumber, and the resulting accelerated particle
spectrum is not a simple power-law in energy (Fig.~\ref{fig:f5}).

As the study here assumed no back-reaction of the waves to the shock
structure or dynamics, the approach is limited to small-amplitude
waves. The test-wave approach, on the other hand, leads to
mathematical singularities at $M \to M_{\rm c} = \sqrt{r}$; at this
limit the antiparallel waves are amplified infinitely. In the analysis
this limitation affects through the simplified calculation of the gas
compression ratio. In more detailed further analysis the effect of the
upstream wave parameters to the calculation of the gas compression
ratio should be taken into account in order to remove the
(non-physical) singularities, as demonstrated by Vainio \&
Schlickeiser (\cite{VS1999}) for nonrelativistic step shocks.

Also the inclusion of accelerating particles will have effects on the
waves; in this treatise the effect of the particles on the waves have
been omitted. However, the resonant wave--particle interactions could
have a damping effect on the turbulence, thus rapidly changing the
transmitted wave distributions, especially when moving downstream away
from the shock. Similarily, our analysis also neglects the wave--wave
interactions in the downstream medium (Vainio \& Spanier
\cite{VainioSpanier2005}).

Finally we merely note that in the present study we have confined the
analysis to strictly parallel shocks. For oblique magnetic field
alignments the treatment of the turbulence transmission becomes very
complicated and is beyond the scope of this study.
%

%
%
%
%

\section{Conclusions}

We have calculated the transmission coefficients for Alvf\'en (test) waves
with short wavelengths through a modified shock with background plasma
properties changing slowly compared to the wave length scales.

While the transmission trough a step shock was found similar
throughout the whole shock speed range (\cite{VVS2003}), and,
additionally, very similar behaviour was observed also for thick
shocks in the nonrelativistic limit (\cite{VV2005AA}), qualitative and
signifigant differences emerge for thick shocks when the speed
increases. As a consequence, from the point of view of the particle
acceleration studies, the compression felt by the particles (i.e. that
of the scattering centres) at a thick relativistic low-$M$ shock was
shown to be weaker than the compression of the background flow. For
slower (from non- to mildly relativistic) shocks the scattering centre
compression ratio was shown to increase, like in step shocks.

%
%
%
%

\begin{acknowledgements}
  The authors thank an anonymous referee for valuable suggestions on
  how to improve the manuscript. J.T. further thanks professors
  Reinhard Schlickeiser and Ian Lerche for constructive criticism and
  inspiring discussions.
\end{acknowledgements}

%
%
%
%

\end{document}